\documentclass[twoside]{spie}  %>>> use for US letter paper
\pdfoutput=1
\usepackage[]{graphicx}
\usepackage[]{amssymb}
\newcommand\micron{\mbox{$\mu$m}}%
\newcommand\arcsec{\mbox{$^{\prime\prime}$}}%
\title{Far-infrared polarimetry from the Stratospheric\\Observatory for Infrared Astronomy}

\author{
  John~E.~Vaillancourt\supit{a},
  David~T.~Chuss\supit{b},
  Richard~M.~Crutcher\supit{c},
  Jessie~L.~Dotson\supit{d},
  C.~Darren~Dowell\supit{a,e},
  D.~Al~Harper\supit{f},
  Roger~H.~Hildebrand\supit{f},
  Terry~J.~Jones\supit{g},
  Alexandre~Lazarian\supit{h},
  Giles~Novak\supit{i}, and
  Michael~W.~Werner\supit{e}
\skiplinehalf
\supit{a}Physics Department, California Institute of Technology, Pasadena, CA 91125, USA; \\
\supit{b}NASA Goddard Space Flight Center, Greenbelt, MD 20771, USA; \\
\supit{c}Department of Astronomy, University of Illinois, Urbana, IL 61801, USA; \\
\supit{d}NASA Ames Research Center, Moffett Field, CA 94035, USA; \\
\supit{e}Jet Propulsion Laboratory, Pasadena, CA 91109, USA; \\
\supit{f}Department of Astronomy \& Astrophysics, University of Chicago, Chicago, IL 60637, USA; \\
\supit{g}Department of Astronomy, University of Minnesota, Minneapolis, MN 55455, USA; \\
\supit{h}Department of Astronomy, University of Wisconsin, Madison, WI 53706, USA; \\
\supit{i}Department of Physics \& Astronomy, Northwestern University, Evanston, IL 60208, USA \\
}

\authorinfo{Further author information: J.E.V.: E-mail: johnv@submm.caltech.edu
}
\begin{document} 
  \maketitle 
%\thispagestyle{myheadings}
%{\bf Draft as of \today.}
%%-----------------------------------------------------------

%The Abstract should concisely summarize the key findings of the
%paper.  It should consist of a single paragraph containing no more
%than 200 words. A list of up to ten keywords should immediately
%follow the Abstract after a blank line.  These keywords will be
%included in a searchable database at SPIE.

\pagestyle{myheadings}
\markboth{Far-infrared polarimetry from SOFIA}{J. E. Vaillancourt et al.}

%%%%%%%%%%%%%%%%%%%%%%%%%%%%%%%%%%%%%%%%%%%%%%%%%%%%%%%%%%%%% 
\begin{abstract}
  Multi-wavelength imaging polarimetry at far-infrared wavelengths has
  proven to be an excellent tool for studying the physical properties
  of dust, molecular clouds, and magnetic fields in the interstellar
  medium.  Although these wavelengths are only observable from
  airborne or space-based platforms, no first-generation instrument
  for the Stratospheric Observatory for Infrared Astronomy (SOFIA) is
  presently designed with polarimetric capabilities.  We study several
  options for upgrading the High-resolution Airborne Wideband Camera
  (HAWC) to a sensitive FIR polarimeter.  HAWC is a $ 12 \times 32$
  pixel bolometer camera designed to cover the 53 -- 215 $\micron$
  spectral range in 4 colors, all at diffraction-limited resolution (5
  -- 21 arcsec). Upgrade options include: (1) an external set of
  optics which modulates the polarization state of the incoming
  radiation before entering the cryostat window; (2) internal
  polarizing optics; and (3) a replacement of the current detector
  array with two state-of-the-art superconducting bolometer arrays, an
  upgrade of the HAWC camera as well as polarimeter.  We discuss a
  range of science studies which will be possible with these upgrades
  including magnetic fields in star-forming regions and galaxies and
  the wavelength-dependence of polarization.
\end{abstract}

%>>>> Include a list of keywords after the abstract 

\keywords{polarimetry, far-infrared astronomy, instrumentation,
  interstellar medium, magnetic fields, dust, airborne astronomy, SOFIA}

%%%%%%%%%%%%%%%%%%%%%%%%%%%%%%%%%%%%%%%%%%%%%%%%%%%%%%%%%%%%%
\section{INTRODUCTION}

Mapping polarimetry at far-infrared/submillimeter wavelengths began on
the Kuiper Airborne Observatory (KAO) more than 20 years ago. Over the
course of the KAO's lifetime polarimetric instruments evolved from
single pixel devices\cite{rhh84} to larger arrays\cite{stokes} and
eventually moved to ground-based observatories\cite{hertz2,scubapol}.
Far-infrared (FIR) and submillimeter observations of dense molecular
clouds, protostellar disks and envelopes, the Galactic center, and one
external galaxy have shown that dust emission from these objects is
polarized (at levels of $\sim 2$ -- 10\%) at almost every
point\cite{archivestokes,archivehertz,li06}. The alignment of dust grains
with the local magnetic field\cite{dg51} allows one to infer the
direction of the magnetic field projected onto the plane of the sky.
This allows studies of magnetic fields' interactions with the local
interstellar medium and their role in star and galaxy
formation\cite{dasth,chuss03,mwf01,m82scuba} as well as studies of
dust and cloud properties.\cite{pspec,lgm97}

As successor to the KAO, the Stratospheric Observatory for Infrared
Astronomy (SOFIA) offers a new frontier for observations in the
infrared-to-submillimeter spectral regime. Unfortunately, no
first-light SOFIA instruments\cite{krabbe02} have existing capability
for accurate polarimetry. Both a far-infrared (\emph{Hale})\cite{hale}
and mid-infrared\cite{packham} polarimeter have been proposed as
future SOFIA instruments.  Here we discuss three options for upgrading
the High-resolution Airborne Wideband Camera (HAWC)\cite{hawc} to a
sensitive FIR polarimeter.

HAWC is a first-light instrument for SOFIA\@. It consists of a $12
\times 32$ pixel bolometer camera designed to cover the 53 -- 215
$\micron$ spectral range in 4 colors, all at diffraction-limited
resolution ($5\arcsec$ -- 21\arcsec).
%As a sensitive camera HAWC will be able to study a wide variety of
%science topics in far-infrared astronomy including star and
%planet formation, the composition of interstellar dust, and conditions
%in external galaxies.
A complete upgrade of the HAWC camera into a polarimeter (which we
call \emph{Hale}) will include replacing the existing 384 pixel array
of silicon bolometers with two superconducting bolometer arrays.  Each
such array would consist of over 5000 pixels and detect orthogonal
polarization components selected by wire grids close to the
detectors. A cold (4\,K) crystal half-wave plate (HWP) upstream from
the grids modulates the incoming radiation. In this way we create a
sensitive, diffraction-limited, 4-color polarimeter.  Additionally,
the dual-polarization measurement and the detector upgrade yield a
photometer with a much larger field-of-view than the original HAWC
instrument.

\begin{figure}[tb]
  \begin{center}
      \includegraphics[width=15cm]{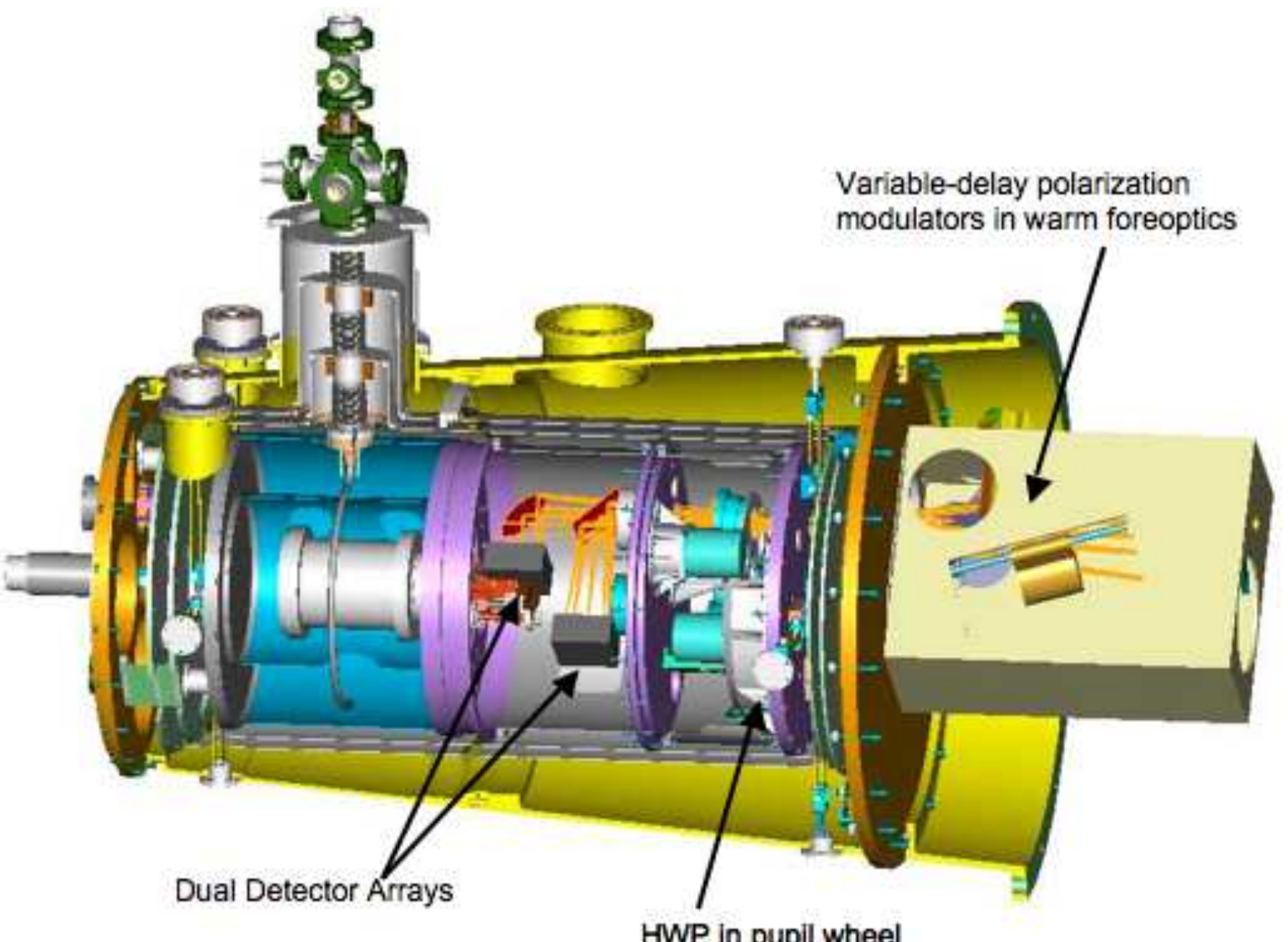}
  \end{center}
  \caption[example] {The HAWC cryostat showing the locations of
    several proposed polarimetric upgrades: dual-polarization detector arrays,
    wave-plates (HWP), and the variable-delay modulator.
    \label{fig:cryostat} }
\end{figure}

%Such an extensive upgrade requires long lead times and perhaps
%significant funding sources.  Therefore, 
We consider two other options (called HAWC-pol) as ``pathfinder''
programs working towards the more complete \emph{Hale} instrument.
These options are attractive as it is feasible that they may be
incorporated into HAWC soon after its first light on SOFIA\@. In both
HAWC-pol options we retain the existing $12\times 32$ silicon
bolometer array and measure only a single component of polarization.
Option (1) places one or more rotating HWPs at the cold HAWC pupil
stop, followed by a polarizing grid.
%The polarization is modulated by rotating the HWP with respect to the
%grid.
%The HWP-grid combination can be switched by rotating a pupil in order
%to select different polarimetric passbands or to completely remove the
%polaring optics from the beam, returning HAWC to its original
%photometric configuration.  
In option (2) we place only a polarizing grid at the pupil and
modulate the polarization outside the cryostat using a variable-delay
polarization modulator, or polarization ``switch''.
%which chooses between orthogonal polarization states. 
Details of the upgrade components are discussed in Sect.\
\ref{sect:instrument}; their locations in the HAWC cryostat are shown
in Fig.~\ref{fig:cryostat}.  The technical specifications for both
\emph{Hale} and HAWC-pol are shown in Table \ref{tbl:specs}.

{\renewcommand{\footnoterule}{\rule{1cm}{0cm}\vspace{-0.7cm}}
%\setlength{\textfloatsep}{0.2in}
%\renewcommand{\footnoterule{\relax}}
%\footnotesize
\begin{table}
\begin{center}
\begin{minipage}{6.6in}
  \caption{HAWC-pol and \emph{Hale} specifications. 
%Values on left (right) sides of each column pertain to HAWC-pol (\emph{Hale}).
}
\begin{tabular}{lcccc}
\hline\hline
Parameter &  Band 1 &  Band 2 & 
             Band 3 &  Band 4 \\ \hline
Central Wavelength ($\micron$) & 
             53 &  89 & 
             155 &  216 \\
%Central Frequency (THz) & 5.7 &  3.4 & 1.9 &  1.4 \\
Bandwidth FWHM ($\Delta\lambda/\lambda$) & 0.16 &  0.19 & 0.22 &  0.21 \\
Pixel Size (arcsec) & 2.3 & 3.5 & 6.0 &  8.0 \\
Resolution FWHM (arcsec) & 5.4 &  9.0 & 16 &  22 \\
Field of view, HAWC-pol (arcmin) &
%\footnote{HAWC-pol: $12\times 32$ pixels; \emph{Hale}: $64\times 80$ pixels}
    $0.5\times 1.2$ & $0.7\times 1.9$ & $1.2\times 3.2$ & $1.6\times 4.3$ \\
Field of view, \emph{Hale} (arcmin) &
%\footnote{HAWC-pol: $12\times 32$ pixels; \emph{Hale}: $64\times 80$ pixels}
   $2.4\times 3.0$ & $3.7\times 4.7$ & $6.4\times 8.0$ & $8.5\times 8.5$ \\
Background Power (pW/pixel) & 24 &  20 & 20 & 10 \\
NEP\footnote{Noise Equivalent Power: background limited, per pixel}
(fW Hz$^{-1/2}$) & 0.43 &  0.30 & 0.24 &  0.14 \\
%NEFD\footnote{Noise Equivalent Flux Density: background limited, per beam, chopped}
%(Jy s$^{1/2}$) & 1.2 &  0.82 & 0.82 & 0.60 \\
NEFD\footnote{Noise Equivalent Flux Density for HAWC camera: background limited, per beam, chopped}
(Jy s$^{1/2}$) & 0.87 &  0.59 & 0.60 & 0.44 \\
Polarization uncertainty,\footnote{In 5 hours for 5\,Jy source; chopped; assumes 100\% polarization efficiency and 100\% observing efficiency.} HAWC-pol (\%) & 0.26 & 0.18 & 0.18 &  0.13 \\
Polarization uncertainty,$^c$ \emph{Hale} (\%) & 
     0.18 & 0.13 & 0.13 & 0.09 \\
Position angle uncertainty ($P=3\%$),$^c$ HAWC-pol (degrees) &
     2.5 & 1.7 & 1.7 &  1.3 \\
Position angle uncertainty ($P=3\%$),$^c$ \emph{Hale} (degrees) &
     1.7 & 1.2 & 1.2 & 0.9 \\
\hline
\label{tbl:specs}
\end{tabular}
\end{minipage}
\end{center}
\vskip 0ex
\end{table}
}

\section{SCIENTIFIC GOALS} \label{sect:science}

\subsection{Turbulent Star Formation} \label{sect:bfields}
%\subsubsection{Turbulent Star-Formation}

The paradigm for star formation, once seen as depending on slow
diffusion of magnetic fields out of cloud cores, has shifted to a
violent one,\cite{mckee07} the dynamics of which are governed by magnetic
compressible turbulence.\cite{biskamp03,cho05}

A unique view of the turbulent field is provided by polarimetry of
emission from magnetically aligned grains. Problems of interpreting
the statistics of 2D polarization maps in terms of the underlying
statistics of the 3D magnetic fields are becoming tractable with
modern theoretical work on radiative transfer, grain alignment, and
turbulence, even in the case of inhomogeneous clouds containing dense
clumps and clusters of embedded stars.\cite{bethell07} The problem of
field bending due to large-scale non-turbulent effects is being
addressed by choosing reference systems along the local projected
direction of the field rather than the mean field for the whole cloud.

When maps of polarized emission can be made with SOFIA's resolution,
dynamic range, and spectral coverage, it will be feasible, not only to
test predictions of turbulent vs.\ static star formation, but also to
determine characteristics of the magnetic fields such as the ratio of
fluctuating to mean components, the energy spectrum, and the
distribution of magnetic field strengths.

%\begin{figure}
%  \begin{minipage}[b]{\textwidth}
%  \parbox[t][10cm][c]{0.5\textwidth}{
%   \includegraphics[width=7cm]{omc1_sharp2007a.eps}
%   } \hfill
%  \parbox[t][5cm][t]{0.50\linewidth}{
%    \caption[example] {Polarization $E$-vectors of the Orion Molecular
%      Cloud (OMC-1) from the SHARP instrument at the Caltech
%      Submillimeter Observatory.  Red and blue vectors represent
%      measurements at 350 and 450 $\micron$, respectively (ref?).  Total flux
%      measurements (grayscale) were taken with the SHARC-II
%      camera (ref?). A 4\% polarization scale is shown in lower left and
%      the approximate $10\arcsec$ beam in the lower right.
%      \label{omc1}}
%}    \end{minipage}
%\end{figure}

\subsection{Dust Grains and the Polarization Spectrum} \label{sect:pspec}

Recent years have been marked by substantial progress in the
understanding of grain alignment\cite{lazarian07}. Research on
radiative torques\cite{dolginov76,draine96,draine97,lh07} and grain
dynamics\cite{spitzer79,purcell79,lazarian94,lazrob97,ld99a,ld99b,weingartner03,hoang07}
have made it possible to explain the otherwise perplexing results of
FIR polarization from starless cores\cite{ward00} and to make quantitative
predictions of grain alignment along lines of sight through dense
clouds.

\begin{figure}[tb]
%  \begin{center}
%    \begin{tabular}{c}
      \includegraphics[width=8.4cm]{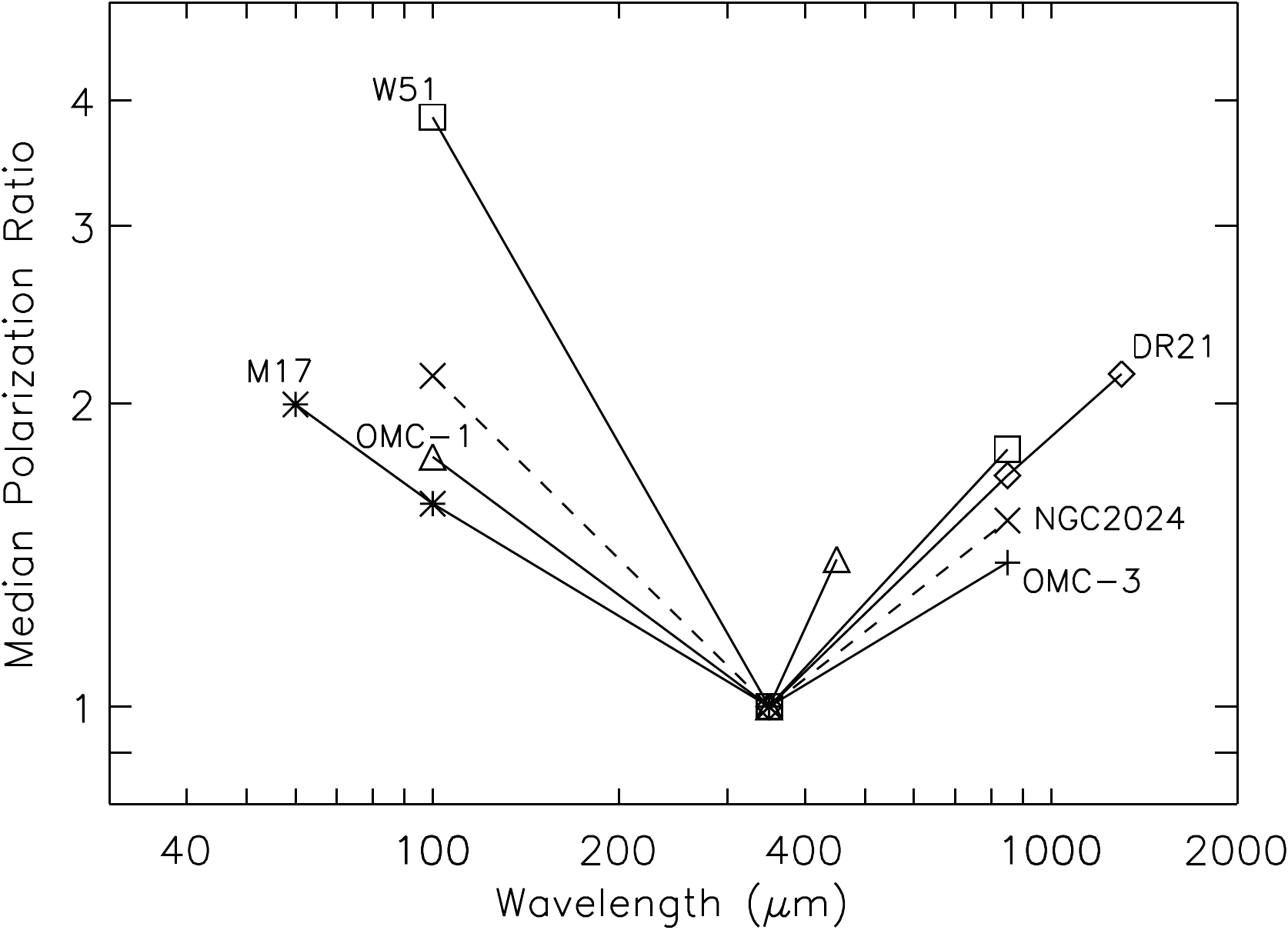}
      \includegraphics[width=8.4cm]{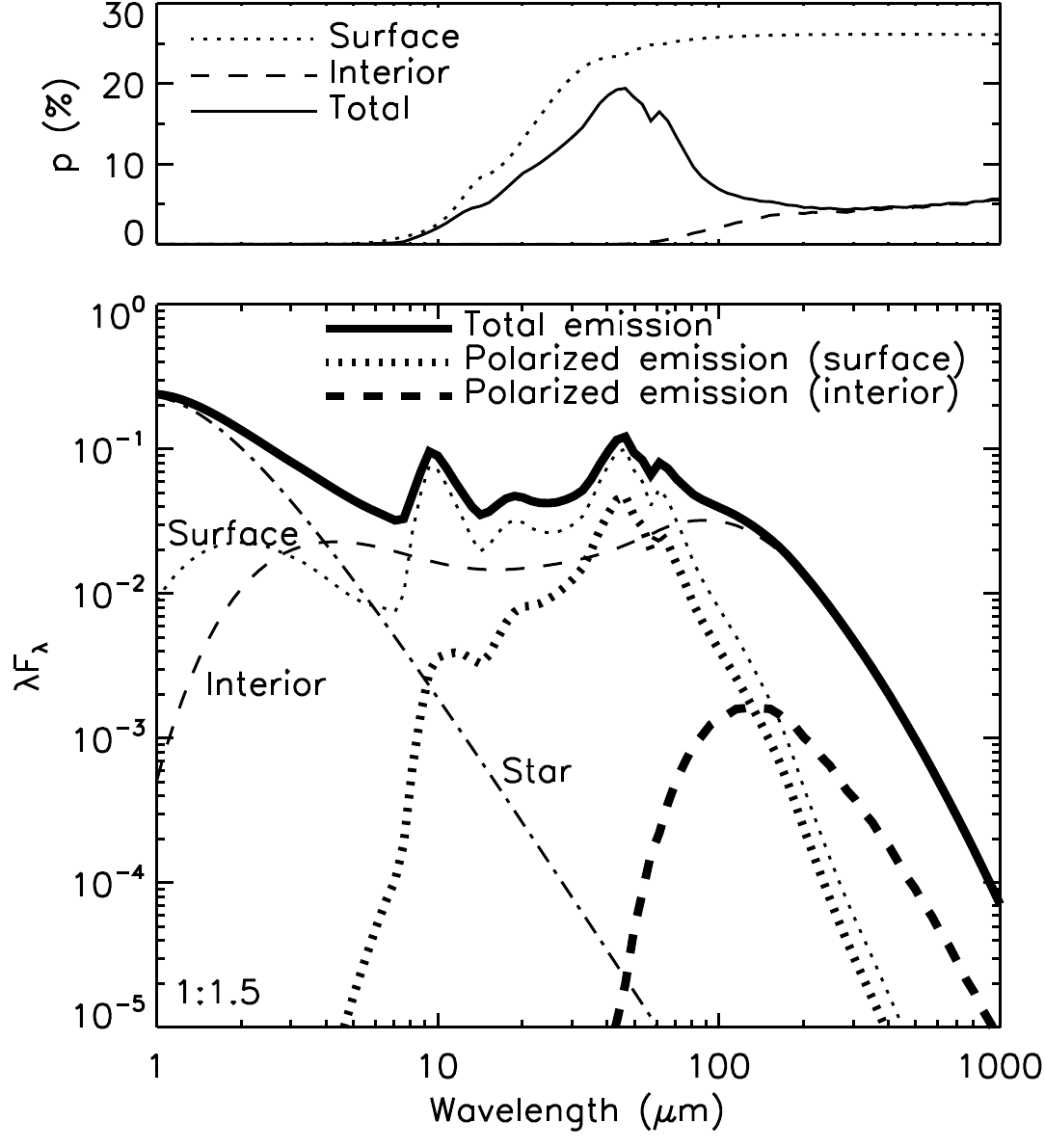}
%    \end{tabular}
%  \end{center}
  \caption{Left: Measured polarization spectra of several molecular
    cloud envelopes, each cloud is normalized at
    $350\,\micron$\cite{paris,sharpomc1}.  Right: Predicted spectral
    energy distribution (bottom) and polarization spectrum (top) for
    protoplanetary disk emission.\cite{cho2007} Note that the peak in
    the polarization occurs at 45\,\micron, corresponding
    approximately to the shortest HAWC passband.}
    \label{fig:pspec}
%}
\end{figure}

The anisotropic radiation responsible for aligning grains is
ubiquitous in astrophysics.  Thus polarized emission from dust grains
can trace magnetic fields in various astrophysical
environments. (While radiative torques provide the aligning mechanism,
the alignment occurs with respect to the local magnetic field.)
Features in the measured polarization spectra of molecular clouds
(Fig.~\ref{fig:pspec}) have been attributed to varying dust
temperatures and polarizing efficiencies along the
line-of-sight\cite{pspec,mythesis,paris}. The alignment of grains,
their temperature, and their emission properties depend on the
radiation field to which they are exposed. As a result,
multi-wavelength polarimetry can be used for detailed studies of
magnetic fields even within unresolved objects
(e.g.\ Sect.~\ref{sect:tts}; Fig.~\ref{fig:pspec}).

A polarimeter on SOFIA will make it possible to pursue these studies
with more passbands, more pixels, and improved spatial resolution.

\subsection{Polarization spectra of protoplanetary disks} \label{sect:tts}

By the time a young stellar object reaches the Class II stage, a.k.a.\
T-Tauri star (TTS) stage, the star has acquired approximately its
final mass, but a residual disk still contains diffuse matter that
will presumably be incorporated into planets.  According to some
current theories of planet formation\cite{dominik07}, the first step
is for the grains in such a protoplanetary disk to coagulate, forming
particles with sizes of order one millimeter or larger.  Observational
evidence of such ``large grains'' has been obtained via modeling of
the submillimeter-millimeter spectral energy distributions of TTS
disks.\cite{beckwith91,dalessio01,rodmann06,draine06} If these models
are correct, then the optical depths for absorption and scattering
will approach or exceed unity across much of the FIR-to-millimeter
spectrum.\cite{dalessio01} In this case, the polarization of the disk
thermal emission will likely contain contributions from selective
absorption and scattering.\cite{krejny06} Although this complicates
the interpretation, it also presents us with a new opportunity:
Because the cross-sections for absorption and scattering, and thus the
polarization characteristics, depend strongly on grain size for sizes
comparable to the wavelength, we may be able to confirm the existence
of millimeter-sized grains and further constrain the grain size via
polarimetry of the disk emission.

Polarimetry of protoplanetary disk emission has been sparse and the
modeling is at an early stage, but with the advent of SOFIA,
ALMA\footnote{{http://www.alma.nrao.edu/}}, and EVLA\cite{napier06} it
will become possible to carry out sensitive polarimetry across the
FIR-to-millimeter wavebands for a sample of bright TTS disks. A recent
model\cite{cho2007} shows that the polarization characteristics depend
strongly on grain size (Fig.\ \ref{fig:pspec}).  Initial polarimetric
observations of TTS disk emission, obtained at the James Clerk Maxwell
Telescope\cite{tamura99} and at the Caltech Submillimeter Observatory
(M. Krejny, priv.\ comm.) suggest polarization levels of one to
several percent.  Four-color polarimetry with \emph{Hale} or HAWC-pol
will place constraints on grain size and provide information on the
magnetic fields (the latter of which may influence disk
evolution\cite{balbus91}).

\subsection{Connecting Large- and Small-Scale Interstellar Fields}

The role played by magnetic fields in star formation has remained
uncertain. At one extreme of the theories of star formation is that
magnetic fields control the formation and evolution of the molecular
clouds from which stars form, including the formation of cores and
their gravitational collapse to form protostars.\cite{MC99} The other
extreme is that magnetic fields are unimportant, with molecular clouds
forming at the intersection of turbulent supersonic flows in the
interstellar medium.\cite{MK04} In spite of considerable effort to
observe magnetic fields in molecular clouds in order to resolve this
uncertainty, the issue remains controversial. SOFIA with \emph{Hale}
or HAWC-pol will provide a crucial link between large- and small-scale
data on interstellar magnetic fields that will be essential to
understand the role of magnetic fields in star formation.

On the largest scales, observations of stellar
polarization\cite{H00,PM07} and Planck\cite{planck} polarization maps
at $5^\prime$ resolution enable mapping of the Galactic to molecular
cloud scale structure of the interstellar magnetic field. HAWC-pol
will make it possible to study structure of cores relative to
large-scale magnetic fields. Some predicted phenomena that require the
correlation of the large- and small-scale maps of magnetic field
structures include collapse of mass along field lines to form cores
flattened along field lines, hourglass morphology fields in cores with
the core fields connecting to the larger scale molecular cloud fields,
magnetic braking of cores that will twist the fields as angular
momentum is transferred outward from cores to envelopes, and bipolar
outflows from protostars with magnetic fields being parallel to the
outflows.

On the smallest scales ($<$ few arcseconds), ALMA will image
protostars, protoplanetary disks, and the inner parts of protostellar
outflows. In spite of the much larger telescope collecting area of
ALMA, the shorter wavelengths and much higher bandwidths observed by
HAWC mean that HAWC-pol will have comparable sensitivity to ALMA for
polarization mapping of extended emission. Theoretical models predict
the morphology of the connection of magnetic fields in these very
small scale phenomena to the larger scale fields in the cores from
which protostars form. Combining ALMA and HAWC-pol maps will probe the
full spatial structure of these phenomena and allow testing of the
star formation theoretical results.

\subsection{The Galactic Center and External Galaxies} \label{sect:external}

\begin{figure}[tb]
%  \begin{minipage}[b]{\textwidth}
%    \parbox[t][8.5cm][c]{0.5\textwidth}{
  \begin{center}
      \includegraphics[width=8.5cm]{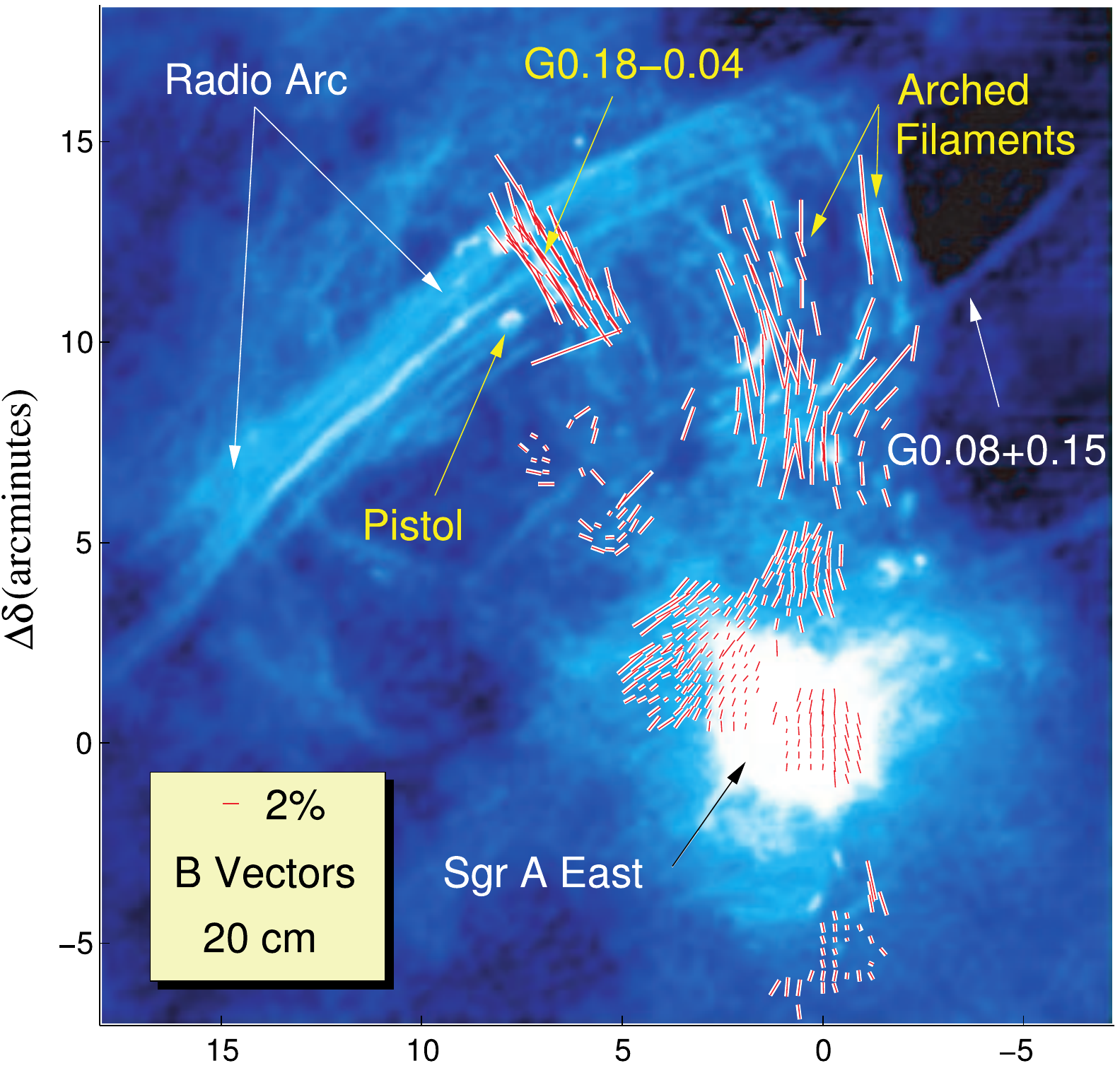}
%      \vskip -2ex
  \end{center}
%    } \hspace{2em} \hfill
% \parbox[t][3.5cm][t]{0.45\linewidth}{ 
  \caption{Polarization and radio map of the Galactic
    center\cite{chuss03}. Inferred magnetic field vectors from
    FIR/submillimeter polarimetry are superposed on a 20\,cm continuum
    VLA image.  The $100\,\micron$ vectors appear to trace the thermal arched
    filaments while the magnetic field in the
    molecular cloud associated with G0.18-0.04 ($350\,\micron$
    polarimetry) is perpendicular to the field traced by the
    non-thermal filaments of the Radio Arc.}
      \label{fig:gc}
%}    \end{minipage}
  \end{figure}

Our proximity to the Galactic Center provides an opportunity to study
the physics of galactic nuclei in great detail. There is evidence that
magnetic fields play an important role in the dynamics of the Galactic
center region as evidenced by the prominent Radio Arc\cite{yz84} and
other non-thermal filaments\cite{yz04,nord04} that exist in this
region.  These filaments are thought to be the result of relativistic
electrons spiraling along lines of magnetic flux, and though the
brightest are oriented perpendicular to the plane, more recently
discovered filaments have been found with other orientations.
Submillimeter polarimetry has indicated that the large-scale magnetic
field in the molecular gas is generally parallel to the
plane\cite{novak03} in stark contrast to the brightest radio
filaments. However, on smaller scales the field has been found to be
more complex (Fig.\ \ref{fig:gc}).\cite{chuss03}

%\begin{figure}
%  \begin{minipage}[b]{\textwidth}
%  \parbox[t][10cm][c]{0.5\textwidth}{
%   \includegraphics[width=7cm]{omc1_sharp2007a.eps}
%   } \hfill
%  \parbox[t][5cm][t]{0.50\linewidth}{
%    \caption[example] {Polarization $E$-vectors of the Orion Molecular
%      Cloud (OMC-1) from the SHARP instrument at the Caltech
%      Submillimeter Observatory.  Red and blue vectors represent
%      measurements at 350 and 450 $\micron$, respectively (ref?).  Total flux
%      measurements (grayscale) were taken with the SHARC-II
%      camera (ref?). A 4\% polarization scale is shown in lower left and
%      the approximate $10\arcsec$ beam in the lower right.
%      \label{omc1}}
%}    \end{minipage}
%\end{figure} 

Key outstanding questions concerning the magnetic fields in the
Galactic center are: (1) What is the geometry of the field, (2) what
is the strength of the field, and (3) how do the radio filaments form?
%\emph{Hale} will address these questions by providing the opportunity
%to study magnetic fields in the warmer dust of the Galactic center at
%higher resolution than has been possible in the past. Its
The multi-frequency capability of a HAWC polarimeter will allow us to
separately measure the field in different components along the
line-of-sight. This is particularly important in the Galactic center,
as the region encompasses a wide range of temperatures. In addition,
the high angular resolution may enable one to probe the interaction
sites between filaments and associated molecular clouds. Such
interactions may provide a mechanism for electron acceleration which is
required for the formation of the filaments. This has the potential
for providing estimates of the local magnetic field strength
and constraining models for filament formation.

External galaxies give us the opportunity to study the magnetic field
geometry of an entire system viewed from the \emph{outside}. Edge on
spiral galaxies in particular will allow us to sample the integrated
interstellar medium along a number of lines of sight, weighted of
course by the distribution of dust, stars, and heating radiation as a
function of galactocentric radius. For many galaxies such as NGC\,891,
NGC\,4565, etc., the edge-on disk is only about one beam wide at
$60\,\micron$ and spans many beam widths across the sky. Comparison of
very active galaxies such as M82, moderately active disks such as in
NGC\,891 and more quiescent disks such as in NGC\,4565 will allow \emph{Hale}
to find and distinguish large-scale structures in the magnetic field
deep within these galaxies. These features could include massive
blowouts, for example, with a projected magnetic field geometry
vertical to the galactic plane, undetectable with optical and
near-infrared polarimetry.

%\newpage
\section{Instrument Design} \label{sect:instrument}

\subsection{Polarization Modulation}

Atmospheric constraints on suborbital FIR/submillimeter
polarimetry have been described in the literature
\cite{clemens90,hildebrand00}.  The two main effects are: 1) the
temporal change of the atmospheric transmission, and 2) ``sky noise'',
which is the spatially and temporally variable emission from the
atmosphere which is not completely eliminated by techniques such as
chopping.  Both effects can be mitigated by a polarimeter which
observes the field of view in two polarizations simultaneously
\cite{hildebrand00}.  However, it is often not practical to implement
the dual-polarization approach in an existing instrument design.  An
alternate solution is to perform polarization modulation quickly, so
as to ``freeze" the atmosphere \cite{clemens90,siringo04}.  Both the
dual-polarization and rapid-polarization-modulation techniques work
because atmospheric emission is intrinsically unpolarized.

For HAWC-pol, we are considering two approaches to rapid polarization
modulation -- a variable-delay polarization modulator (VPM), and a
continuously-spinning half-wave plate (HWP\@).  Both devices are
considered in more detail in the following sections.  However, several
aspects of the design are common.

In our observing plan, the chopping secondary mirror of SOFIA is the
fastest modulation, and the VPM or HWP is a slower modulation.
Secondary chopping is ideally done at a suitably high frequency where
the noise of the difference is dominated by white
noise from the photon arrival statistics\cite{duncan95}.  The typical
atmospheric noise spectrum for SOFIA FIR observations is not
yet known, but the mechanical capabilities support chopping at 5 -- 20
Hz, which we expect will be fast enough in most conditions.

Although the atmospheric sky noise is intrinsically unpolarized, it
will become partially polarized by non-normal reflections in the
optics, in particular the dichroic tertiary mirror.  Past experience
with this type of mirror \cite{archivestokes} indicates that dichroic
mirror polarization will be $\sim1$\%.  Therefore, the sky noise will
be registered as a polarization noise at a level $\sim 100 \times$
reduced.  Since the polarization modulation is at a lower frequency,
the frequency spectrum of the atmospheric noise matters.  If the noise
spectrum is proportional to $1/f^2$ (in amplitude units), then the
polarization modulation needs to be at a frequency at least as large
as $\sim 0.1\times$ the requirement for the chopping.  For HAWC-pol,
we plan on polarization modulation at $\sim 1$\,Hz with either a
square-wave profile (VPM) or sinusoidal profile (HWP or VPM).
%A straightforward calculation relating sky emission noise to sky
%transmission indicates that this rotation speed will also meet the
%requirement to avoid large ($\sim 1$\%) transmission fluctuations on 1
%second timescales.

% 100 microns, 10" beam, 50% transmission, 250 K:  47147 Jy
% detectable sky noise at 10 Hz:  1 Jy/sqrt(Hz) * sqrt(10 Hz) = 3 Jy ->
% 300 Jy noise at 1 Hz:  47447 Jy -> 49.7% transmission -> 0.6% change

Since all of the radiation from the atmosphere is partially polarized
by the dichroic mirror, continuous polarization modulation will
produce a large, nearly constant periodic signal on the detectors
equivalent to a $\sim 1$\,K (Rayleigh-Jeans) load.  The detector
readout must have a dynamic range which accepts this signal, and the
linearity of the detector over this range must be understood.  Since
the polarization modulation is used with faster spatial modulation,
the celestial polarization signal will appear at $\sim 1$\,Hz
``sidebands'' of the chopping frequency and its harmonics; the
detector readout must have good noise performance in this frequency
range.  All of these requirements are met by the HAWC system.  The
large 1\,Hz atmospheric signal can be filtered out in an early data
analysis step.

\subsection{A Continuously-Spinning Half-Wave Plate}

In this section, we describe an approach to HAWC polarimetry using a
uniformly rotating birefringent quartz HWP\@.  Cryogenic crystal
quartz has well characterized o-ray and e-ray indices of refraction,
and thin waveplates should have loss $< 5$\% over most of the FIR.
\cite{nordh87,brehat97} Although waveplates can be stacked to achieve
near-achromaticity over an octave of wavelength\cite{scubapol}, the
HAWC filter bands cover two octaves.  We expect that at least two
HWPs will be required and perhaps four.

% Brehat & Wyncke:
%  50 um:  delta n = 0.065 -> d = 385 um
%          k = 2.5e-4 -> alpha = 0.63 cm-1 -> loss ~ 2.4%

We have identified the HAWC pupil wheel as an ideal location to locate
polarimeter HWPs.  An existing mechanism provides infrequent selection
of up to eight pupil apertures for diagnostic purposes.  In practice,
no more than four of these apertures are needed for the
camera\cite{hawc}, leaving four positions available for HWPs.  The
pupil wheel is located at an image of the SOFIA primary mirror, which
is an optimum location for a HWP.\cite{gonatas89}  The pupil image
has a diameter of 38\,mm, a manageable size for quartz elements.  A
wire-grid polarizer will be located behind each HWP to
provide polarization detection.  The polarimeter is removed from the
optical path by rotating the pupil wheel to select the camera-mode
pupil aperture.

For 1\,Hz polarization modulation, the HWP must rotate at 15 rpm.
Even at this low speed, induced vibration and microphonic response of
the high-impedance bolometer detectors is a concern.  We will avoid
ball bearings and gear trains in the design and will instead use
jewel bearings and three rollers to suspend each half-wave plate and a
motor which directly torques permanent magnets attached to the
waveplate.  We expect dissipation in the bearings to be under 1\,mW,
which can be carried away through the pupil wheel.  An optical encoder
will monitor the HWP position.  Implementing this polarimeter solution
is straightforward and will require reconstruction of the pupil wheel,
the plate which holds it, and the Geneva mechanism which defines the
eight positions.

\subsection{The Variable-delay Polarization Modulator (VPM)} \label{sect:vpm}

\begin{figure}
%\begin{center}
%\includegraphics[height=1.34in]{VPMfig.eps}
%\includegraphics[height=1.34in]{diagram.eps}
\includegraphics[height=1.34in]{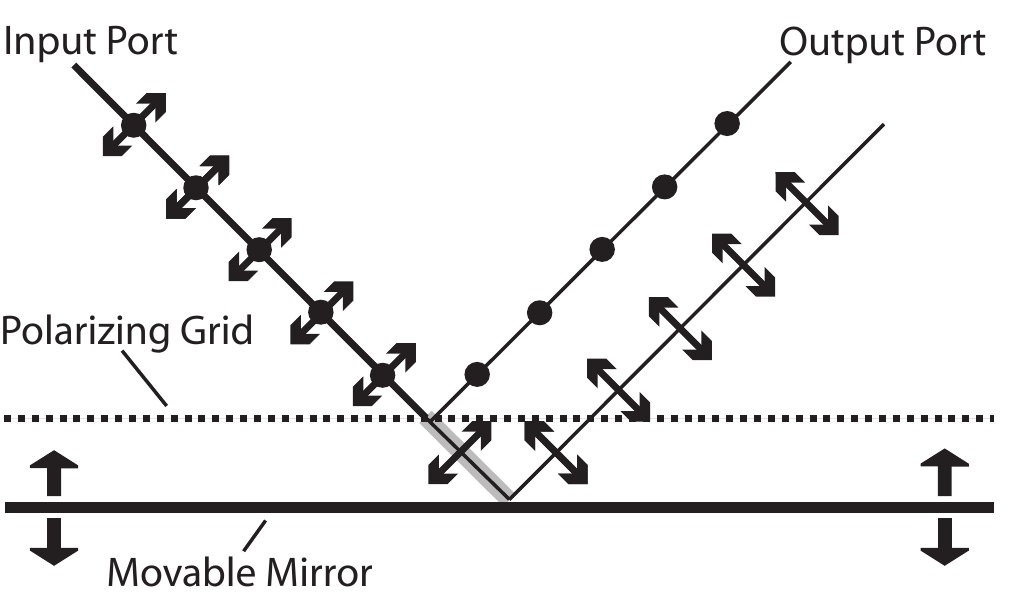}
\includegraphics[height=1.34in]{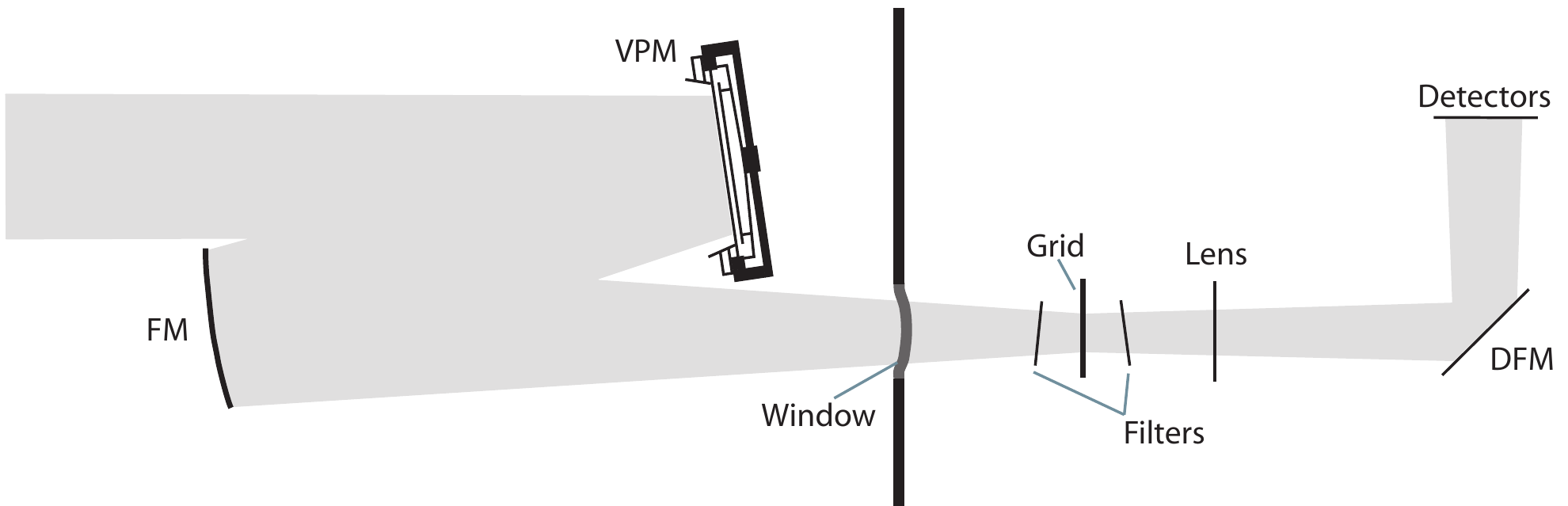}
%\end{center}
\caption{Left: A schematic for the Variable-delay Polarization
  Modulator (VPM\@). A controlled, variable delay (shaded line) can be
  introduced between two orthogonal linear polarizations. Right: The
  current HAWC fore-optics can be modified for the polarimeter. The
  optics to the left of the window are at the ambient
  temperature; these include the VPM and the Focusing Mirror
  (FM\@). The optics to the right of the window are all $\sim 4$\,K; these
  include the analyzing grid, filters, lens, detector folding mirror
  (DFM), and the detectors themselves.}
\label{fig:VPM}
\end{figure}

%Polarization modulation is the encoding of the polarization signal via
%the systematic transformation of one polarization state into another
%for subsequent demodulation and detection.  The most common way of
%doing this is by introducing a phase delay between two orthogonal
%polarizations.

In the case of the Variable-delay Polarization Modulator (VPM),
polarization is modulated by allowing a controlled phase delay to be
introduced between two orthogonal linear polarizations.  The basic
principle is shown in Fig.\ \ref{fig:VPM}, and consists of a
grid/mirror pair.  Incident partially-polarized radiation comes in
from the left. One component of polarization is reflected from the
front of a grid. The other passes through the grid to a parallel
mirror and is reflected back through the grid where the two beams are
recombined. The component transmitted by the grid experiences an
additional path length (which is shaded in the figure).  As the
grid-mirror separation is modulated, the polarization measured by
polarization-sensitive detectors changes in a predictable way
\cite{Chuss06}.  Various instruments have utilized devices with this
architecture and have referred to them by various names
\cite{Manabe03,Battistelli02,Chuss04,Chuss06,Houde01}.  The
nomenclature ``VPM'' used here emphasizes the functionality as a
polarization modulator and that the degree of freedom employed is that
of a variable phase delay between linear polarizations.

The polarization modulation for HAWC-pol must be flexible enough to
accommodate all 4 HAWC passbands.  As long as the grid is sufficiently
fine for the $53\,\micron$ band, we only need to change the grid-mirror
separations to optimize the VPM for polarization modulation across the
different bands.

By making two changes to the current HAWC configuration, we can use a
single VPM to measure Stokes Q and U for the four HAWC passbands: 1)
Replace the warm folding flat mirror with a VPM that is mounted on a
rotatable stage, and 2) Add two analyzer grids to the open slots on
the aperture (pupil) wheel that have a relative orientation of
45$^\circ$.  Figure \ref{fig:VPM} also shows the warm and cold optics
for HAWC with the proposed modifications for a VPM-based polarimeter.
 
The polarimeter will have two modes, one to measure Stokes Q in the
instrument coordinate system and the other to measure Stokes U\@. These
will be differentiated by use of the two different analyzer grids that
are oriented at a 45$^\circ$ with respect to one another.  The VPM is
placed on a rotator such that the relative angle of the VPM wires with
respect to whichever analyzer grid is used is always 45$^\circ$. In
this way the measured signal on the detectors in each of the modes is
\begin{eqnarray}
S_Q = I + Q \cos(\Delta \phi) + V\sin(\Delta \phi), \\
S_U = I + U\cos(\Delta \phi) + V\sin(\Delta \phi),
\end{eqnarray}
where $\Delta \phi$ is the phase delay introduced by the VPM\@.  This
can be scaled to any HAWC frequency by changing the amplitude of the
grid-mirror separation. By using half- and full-wave delays for a
given band, we can use the VPM as a polarization ``switch'' to rapidly
change the polarization state to which the detectors are sensitive
from $I+Q\rightarrow I-Q$ or $I+U\rightarrow I-U$ depending upon the
mode of operation. Table \ref{tab:HAWCpol} illustrates the
polarization measurement strategy.

\begin{table}[tb]
%\vskip 1ex
\begin{center}
\caption{Measuring Polarization with a VPM}
\begin{tabular}{c c c}
\hline\hline
VPM Rotation Angle & $\Delta \phi$ & Polarization State\\
(degrees) & (radians) & \\  \hline
0	&  	$2\pi$	& 	$I+Q$\\
0	&	$\pi$		&	$I-Q$\\
$45$ & $2\pi$	& 	$I+U$\\
$45$ & $\pi$	& 	$I-U$\\ \hline
\label{tab:HAWCpol}
\end{tabular}
\end{center}
\vskip -6ex
\end{table}

%The VPM will be modulated quickly enough such that Stokes I is
%constant over the modulation period. For example, if the secondary
%mirror is chopping at 10\,Hz, we can modulate the polarization at 2--3
%Hz. This will enable us to use the VPM to mimic the effect of
%measuring both polarizations at once.

\subsection{New Technology Large-format Detector Arrays} \label{sect:bolometers}

To enable revolutionary FIR polarimetry, \emph{Hale} will
require simultaneous detection of two orthogonal linear
polarizations. To accomplish this, we plan to implement two focal planes
with 5,120 detectors each, using Goddard Space Flight Center's
Backshort-Under-Grid (BUG) technology\cite{allen06}.  As in the
SCUBA-2 design\cite{holland06}, four such $32 \times 40$ subarrays
will be tiled to produce each 5,120 element array.
 
\begin{figure}
  \begin{center}
    \includegraphics[width=15.9cm]{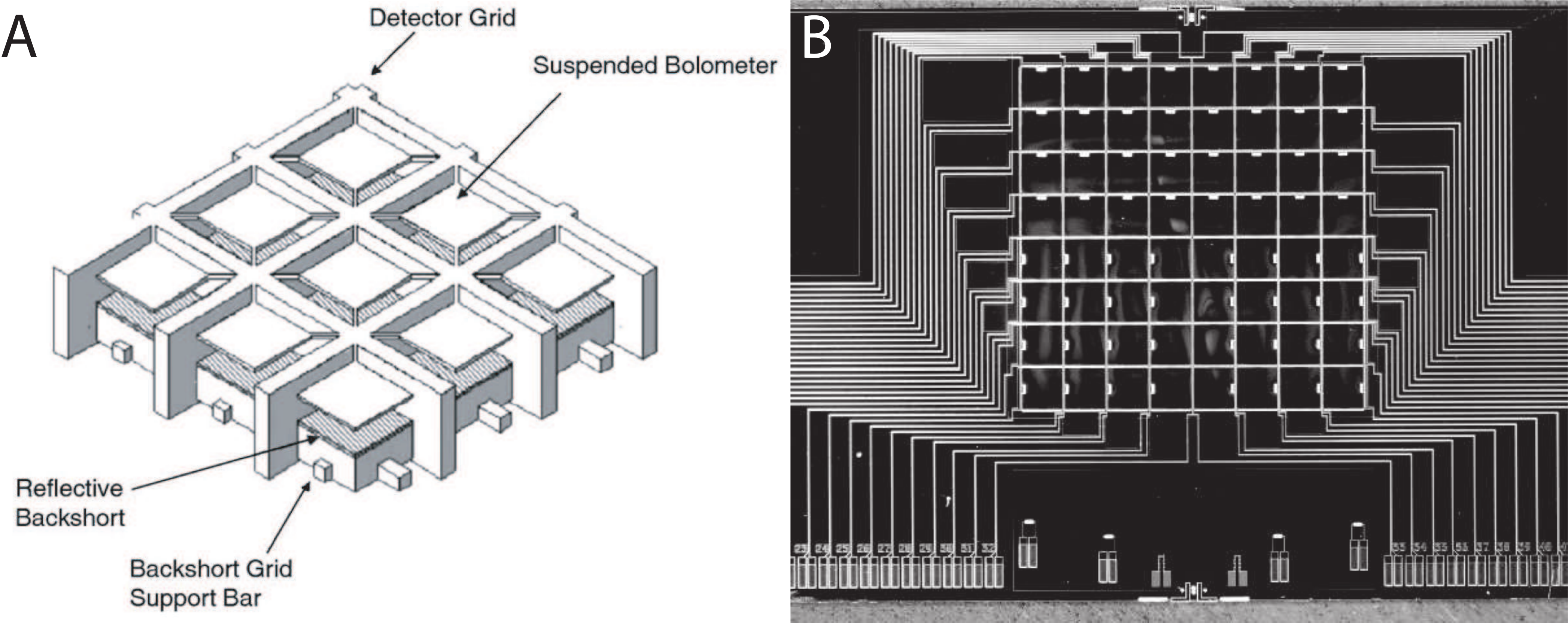}
  \end{center}
  \vskip -4ex
    \caption{a) The Backshort-Under-Grid (BUGs) architecture is shown
  schematically. The absorber/detector membranes are suspended in a
  frame of silicon. Electrical and thermal connections to the frame
  are made over thin silicon legs. The backshort/termination assembly
  fits into keyed structures in the detector grid frame. Electrical
  connections are made to the SQUID multiplexer by wrap-around vias
  that are made along the grid frame. b) A prototype $8 \times 8$
  detector frame is shown. For this prototype, the detector leads are
  fanned out in the plane of the detectors rather than wrapped around
  to the back of the assembly as they will be for the $32 \times 40$
  arrays.\cite{allen06} }
\label{fig:BUGdia}
\end{figure}

The BUG architecture is a two-piece, keyed assembly in which a
two-dimensional array of backshorts are mounted under a matching array
of suspended membranes.  (It should be noted here that for
\emph{Hale}, the term ``backshort'' is a misnomer because in order to
enable good performance over all four passbands, we will actually
employ a ``back termination.'').  Transition-Edge Sensors (TES) are
fabricated on the membranes and the leads are ``wrapped around'' the
frame to the back of the device at which point electrical connections
to the multiplexer can be made via bump-bonding.  The pixel size in
this architecture is 1.135\,mm so that an array of BUGs can be mounted
to the SCUBA-2 32$\times$40 two-dimensional time-domain SQUID
multiplexer developed by NIST\@.\cite{irwin04} Figure \ref{fig:BUGdia}a
shows a schematic of the BUG architecture.

The suspended silicon membranes on the front piece of the BUG assembly
will be ion-implanted to achieve a surface impedance of $157\,\Omega$
per square.  This process has been demonstrated for the Atacama
Cosmology Telescope (ACT)\cite{niemack06} detectors.
%These detectors have similar surface impedance requirements to
%\emph{Hale}.
The back short assembly, which for single-band implementations is made
reflective, will be coated with an absorbing film to prevent unwanted
resonances. This absorbing strategy is similar to that used on HAWC,
and will result in a 50\,\% efficiency across all four passbands.
As in the ACT detectors, a TES consisting of a molybdenum-gold
bilayer with noise-reducing normal metal bars will be fabricated at
the edge of each pixel. 

NASA/Goddard has developed an $8 \times 8$ pixel version of this array
as a step towards the realization of kilopixel arrays. One such array
is shown in Fig.\ \ref{fig:BUGdia}b.  In this device, the TES leads
are routed on the top layer of the device to bond pads at the edge of
the structure. Multiplexing is done external to the detector chip.  An
$8 \times 16$ version with 2\,mm pixels has been fabricated for the
GISMO (Goddard-IRAM Superconducting (2)-Millimeter Observer)
Instrument.\cite{staguhn06}

%\begin{figure*}
%%\begin{center}
%\includegraphics[width=17cm]{diagram.eps}
%%\end{center}
%\caption{The current HAWC fore-optics can be modified for the
%  polarimeter. The optics to the left of the window are at the ambient
%  temperature. These include the Variable-delay Polarization Modulator
%  (VPM) and the Focusing Mirror (FM). The optics to the right of the
%  window are all at 4.2\,K. These include the analyzing grid, filters,
%  lens, detector folding mirror (DFM), and the detectors themselves.}
%\label{fig:diagram}
%\end{figure*}

%% Use of [h] in following command forces table to appear "here".
%\begin{table}[h]
%\caption{Fonts sizes to be used for various parts of the manuscript.} 
%\label{tab:fonts}
%\begin{center}       
%\begin{tabular}{|l|l|} %% this creates two columns
%% |l|l| to left justify each column entry
%% |c|c| to center each column entry
%% use of \rule[]{}{} below opens up each row
%\hline
%\rule[-1ex]{0pt}{3.5ex}  Article title & 16 pt., bold, centered  \\
%\hline
%\rule[-1ex]{0pt}{3.5ex}  Author names and affiliations & 12 pt., normal, centered   \\
%\hline 
%\end{tabular}
%\end{center}
%\end{table} 
 
%\appendix    %>>>> this command starts appendixes

\acknowledgments     %>>>> equivalent to \section*{ACKNOWLEDGMENTS}       
%In the Acknowledgments section, appearing just before the References, the authors may credit others for their guidance or help.  Also, funding sources may be stated.  The Acknowledgments section does not have a section number. 
We would like to thank Fabian Heitsch, Jungyeon Cho, Diego
Falceta-Gon\c{c}alves, Megan Krejny, Jesse Wirth, Harvey Moseley, and
Leslie Looney for useful discussions regarding the material in this
manuscript.  This work has been partially supported by NSF grants
0505124 to the University of Chicago, AST-0540882 to the California
Institute of Technology, and AST-0505230 to Northwestern University.
A.L. acknowledges support from the NSF Center for Magnetic
Self-Organization in Laboratory and Astrophysical Plasmas and grant
AST-0507164. R.M.C. acknowledges partial support from NSF grant
AST-0606822.

%%%%% References %%%%%
%The References section does not have a section number.  The references are numbered in the order in which they are cited.
\bibliography{spie2007,hawcpol_hwp}
\bibliographystyle{spiebib}   %>>>> makes bibtex use spiebib.bst

\end{document}